\documentclass[11pt]{article}

\usepackage[final]{acl}

\usepackage{times}
\usepackage{latexsym}

\usepackage[T1]{fontenc}

\usepackage[utf8]{inputenc}
\usepackage{booktabs}
\usepackage{microtype}

\usepackage{inconsolata}

\usepackage{graphicx}
\usepackage{amsmath}
\usepackage{placeins}
\title{CSSG: Measuring Code Similarity with Semantic Graphs}

\author{
  Yiyang Lu\footnotemark[1],
  Jingwen Xu\thanks{Equal contribution.},
  Changze Lv,
  Zisu Huang,
  Zhengkang Guo, \\
  \textbf{
  Zhengyuan Wang,
  Muzhao Tian,
  Xuanjing Huang\footnotemark[2] and
  Xiaoqing Zheng\thanks{Corresponding author.}
  } \\
  College of Computer Science and Artificial Intelligence \\
  Fudan University \\
  \texttt{\{xujw24, yylu24\}@m.fudan.edu.cn} \\ 
  \texttt{\{xqzheng\}@fudan.edu.cn}
}

\begin{document}
\maketitle
\begin{abstract}
Existing code similarity metrics, such as BLEU, CodeBLEU, and TSED, largely rely on surface-level string overlap or abstract syntax tree structures, and often fail to capture deeper semantic relationships between programs.
We propose CSSG (Code Similarity using Semantic Graphs), a novel metric that leverages program dependence graphs to explicitly model control dependencies and variable interactions, providing a semantics-aware representation of code.
Experiments on the CodeContests+ dataset show that CSSG consistently outperforms existing metrics in distinguishing more similar code from less similar code under both monolingual and cross-lingual settings, demonstrating that dependency-aware graph representations offer a more effective alternative to surface-level or syntax-based similarity measures.
\end{abstract}

\section{Introduction and Related Work}
Functional correctness measured by unit tests is a reliable criterion for code evaluation and has been widely adopted in benchmarks such as HumanEval~\cite{chen2021evaluating}, MBPP~\cite{austin2021program}, and MultiPL-E~\cite{cassano2022multipl}. However, it relies on carefully curated test suites and runtime execution. Moreover, in practice, evaluation is often required at intermediate stages of code generation or for partial code fragments that cannot be compiled in isolation. In such settings, functional testing becomes inapplicable.

Code similarity to reference implementations has therefore long been used as a complementary means of assessing code quality~\cite{iyer2018mapping, hendrycksapps2021}. Early approaches adapted techniques from natural language processing. 
BLEU~\cite{papineni2002bleu} and Jaccard similarity~\cite{jaccard1901etude, roy2009comparison} measure surface-level token overlap and are highly sensitive to formatting or syntactic variations. Empirical studies have shown that BLEU correlates poorly, and in some cases even negatively, with functional correctness~\cite{kulal2019spoc, hendrycks2021measuring}. These limitations motivate the development of code-aware similarity metrics that move beyond purely lexical matching and better reflect program structure.
CodeBLEU~\cite{ren2020codebleu} extends $n$-gram matching with syntactic and data-flow components, while TSED~\cite{song2024revisiting} computes edit distance over abstract syntax trees (ASTs) to reflect hierarchical program structure more faithfully. While these methods improve upon lexical similarity, they primarily operate on AST representations, which capture syntactic structure but abstract away critical semantic dependencies that arise during execution.

In particular, AST-based representations do not explicitly model data and control dependencies that govern runtime behavior and execution logic. Program Dependence Graphs (PDGs)~\cite{ferrante1987program} provide a richer semantic abstraction by directly encoding these dependencies, which largely characterize how programs are conditionally executed and how values are propagated across statements, and have been widely used in program analysis. We further illustrate this distinction through a case study in Appendix \ref{sec:appendix_comparison}. Building on this insight, we introduce \textbf{CSSG} (Code Similarity using Semantic Graphs), a metric that computes edit distance over enhanced PDG-based representations to enable fine-grained, semantically grounded code similarity assessment.

We evaluate CSSG alongside several existing metrics on the CodeContests+ dataset~\cite{wang-etal-2025-codecontests}. Our evaluation considers both monolingual and cross-lingual settings and includes comparisons between pairs of correct solutions as well as correct and incorrect implementations. The results show that CSSG more effectively distinguishes correct from incorrect solutions, and we conduct an explicit correlation analysis to account for the observed performance differences among metrics.

\begin{figure*}[t]
    \centering
    \vspace{-5pt}
    \includegraphics[width=\textwidth]{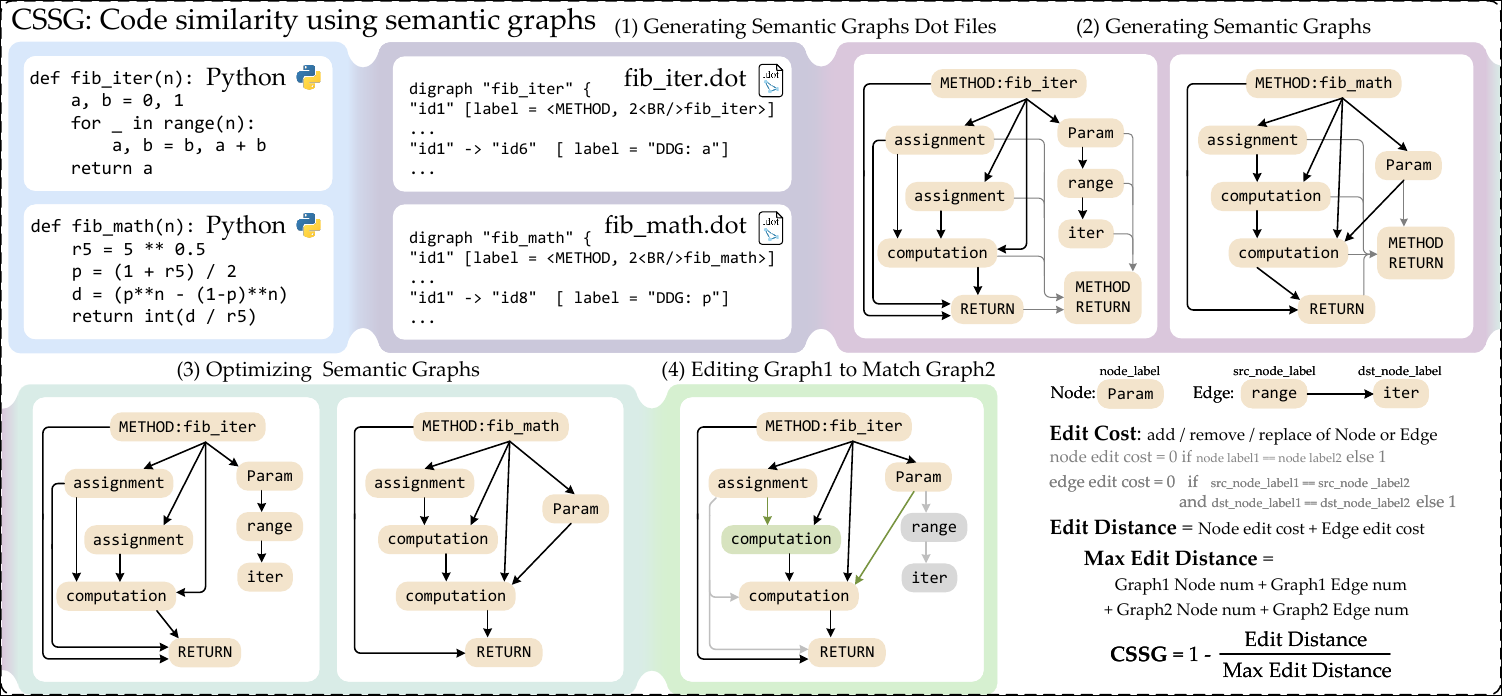}
    \caption{Computation pipeline for semantic graph edit distance. The process consists of four main steps: (1) utilizing Joern to extract function-level PDGs from the code snippet; (2) constructing a unified semantic graph by optimizing nodes and integrating a global root with function call edges; (3) computing the constrained Graph Edit Distancebetween the reference and target graphs; and (4) calculating the final similarity score by normalizing the GED against the maximum edit distance.}
    \label{fig:main}
    \vspace{-5pt}
\end{figure*}

\section{Method}

Given a code snippet, our objective is to compute a semantically grounded similarity score by comparing its program dependence structure with that of a reference implementation. To this end, we propose a graph-based similarity metric built upon an enhanced PDG-based representation and evaluated via constrained graph edit distance.

\subsection{Sematic Graph Construction}

For a given code snippet, we first extract PDGs for all contained functions. Each function-level PDG encodes both data and control dependencies among program statements.

We represent each PDG as a directed labeled graph
\[
G = (N, E, L),
\]
where \( N \) denotes the set of nodes, \( E \subseteq N \times N \) denotes the set of directed edges, and \( L: N \rightarrow \mathcal{L} \) is a node labeling function. Node labels encode semantic categories such as operation type, identifier class, or constant type, while edges represent either data or control dependencies.

To obtain a unified representation at the code-snippet level, we introduce a global root node \( n_g \) and connect it to the entry node of each function-level graph. The resulting integrated graph is denoted as
\[
\tilde{G} = (\tilde{N}, \tilde{E}, \tilde{L}),
\]
where \( \tilde{N} = \bigcup_i N_i \cup \{n_g\} \) and \(
\tilde{E} = \bigcup_i E_i \;\cup\; E_{\text{call}}.
\)
Here, \( E_{\text{call}} \) denotes function call edges that link call-site nodes within a function to the entry nodes of the corresponding callee functions, thereby capturing invocation relationships across functions. This construction lifts function level PDGs to a unified code snippet level semantic graph, enabling holistic comparison while preserving each function’s internal dependence structure.

\subsection{Graph Matching Constraints}

Given two integrated graphs \( \tilde{G}_1 \) and \( \tilde{G}_2 \), we compute their similarity under constrained graph matching. Matching is initiated by fixing the correspondence between the global root nodes of the two graphs, ensuring alignment at the snippet level.

Node matching is subject to the following matching constraints:
\begin{itemize}
    \item \textbf{Function nodes} may be matched only if their function name labels are identical.
    \item \textbf{Non-function nodes} may be matched if their normalized labels are identical. 
\end{itemize}

\subsection{Graph Edit Distance}

Under the matching constraints above, we define the graph edit distance between two integrated graphs \( \tilde{G}_1 \) and \( \tilde{G}_2 \) as the minimum cost required to transform one graph into the other. We consider three types of edit operations: node or edge insertion, deletion, and substitution. All edit operations are assigned equal cost.

Formally, the graph edit distance is defined as
\[
\mathrm{GED}(\tilde{G}_1, \tilde{G}_2) = \min_{\mathcal{O}} \sum_{o \in \mathcal{O}} \mathrm{cost}(o),
\]
where \( \mathcal{O} \) denotes a sequence of valid edit operations transforming \( \tilde{G}_1 \) into \( \tilde{G}_2 \).

\subsection{Similarity Normalization}

To obtain a bounded similarity score, we normalize the edit distance by the theoretical maximum edit distance between the two graphs, defined as
\[
D_{\max} = |\tilde{N}_1| + |\tilde{E}_1| + |\tilde{N}_2| + |\tilde{E}_2|.
\]

The final similarity score is computed as
\[
\mathrm{CSSG}(\tilde{G}_1, \tilde{G}_2) = 1 - \frac{\mathrm{GED}(\tilde{G}_1, \tilde{G}_2)}{D_{\max}},
\]
where a score of \(1\) indicates maximal similarity and \(0\) indicates minimal similarity.

\section{Experiments}

\subsection{General Setups}
The primary objective of this study is to investigate the effectiveness of various code similarity metrics across multiple programming languages. 
Our experiments cover two main tasks. First, we evaluate metrics on monolingual code pairs and analyze correlations among them to reveal their behaviors. Second, since many real-world datasets provide reference solutions in only a specific language, we assess metrics on cross-lingual code pairs to test their ability to capture semantic equivalence across languages.
This design closely mirrors practical scenarios, enabling an analysis of metric strengths and limitations under realistic conditions.
\subsection{Dataset}

A variety of public datasets provide ground truth code and have been widely used in prior work~\cite{yin2018mining, iyer2018mapping, liao2023context, yu2024codereval}. However, these datasets do not fully meet the requirements of our evaluation setting. Existing benchmarks are typically limited to a single programming language~\cite{yin2018mining, iyer2018mapping, liao2023context} or contain only a small number of cases~\cite{yu2024codereval}. More importantly, they generally lack a sufficiently large and diverse collection of both correct and incorrect implementations, which is crucial for evaluating a metric’s ability to distinguish valid solutions from faulty ones.

In our experiments, we use data from CodeContests+~\cite{wang-etal-2025-codecontests} for evaluation. CodeContests+ comprises 11,690 competitive programming problems and over 13 million submissions, including both correct and incorrect solutions. This large-scale collection provides diverse positive and negative examples across various problems and programming languages, making it well-suited for realistic assessment of code similarity metrics.

To ensure computational efficiency and balanced coverage across language pairs, we randomly sample a subset and construct test triplets for each problem in our evaluation. In monolingual tasks, each triplet consists of two correct solutions (\texttt{pos1}, \texttt{pos2}) and one incorrect solution (\texttt{neg}). In cross-lingual tasks, we use one correct solution from the target language (\texttt{pos1}) paired with one correct (\texttt{pos2}) and one incorrect (\texttt{neg}) solution from the source language for each problem.. From each triplet, we derive one positive pair (\texttt{pos1}, \texttt{pos2}) and one negative pair (\texttt{pos1}, \texttt{neg}). The resulting triplet counts are summarized in Table~\ref{tab:sample_stats}.

\begin{table}[htbp]
\centering
\small
\caption{Number of Test Triplets by Language Pair}
\label{tab:sample_stats}
\begin{tabular}{lc}
\toprule
Language Pairs & Number of Triplets \\
\midrule
(C++, C++)       & $10,647$ \\
(Python, Python) & $7,043$ \\
(Java, Java)   & $9,299$ \\
\midrule
\midrule
(C++, Python)    & $7,039$ \\
(C++, Java)      & $9,279$ \\
(Python, C++)   & $7,039$ \\
(Python, Java)   & $6,922$ \\
(Java, C++)      & $9,279$ \\
(Java, Python)   & $6,922$ \\
\bottomrule
\end{tabular}
\end{table}

\section{Results}

\subsection{Monolingual Similarity Results}

To quantify each metric's ability to discriminate between positive and negative code pairs, we report Cohen’s $d$ effect size, which measures the standardized difference between the mean similarity scores of the two groups. By normalizing the separation by pooled variance, Cohen’s $d$ offers a stable measure of discriminative strength that is less sensitive to differences in sample size.

\begin{table}[ht]
\centering
\small
\setlength{\tabcolsep}{3pt}
\caption{Cohen’s d effect sizes of metrics on monolingual code pairs.}
\label{tab:cohen-d-monolingual}
\begin{tabular}{lccccc}
\hline
Languages & BLEU & Jaccard & CodeBLEU & TSED & CSSG	\\
\hline
C++ & 0.122 & 0.102 &  0.084 & 0.179 & \textbf{0.223} \\
Python & 0.027 & 0.055 & 0.086 & 0.211 & \textbf{0.213} \\
Java & -0.028 & -0.024 & 0.018 & -0.020 & \textbf{0.046} \\
\hline
Average & 0.040 & 0.044 & 0.063 & 0.123 & \textbf{0.161} \\
\hline
\end{tabular}
\end{table}

\begin{figure}[t]
    \centering
    \vspace{-5pt}
    \includegraphics[width=\linewidth]{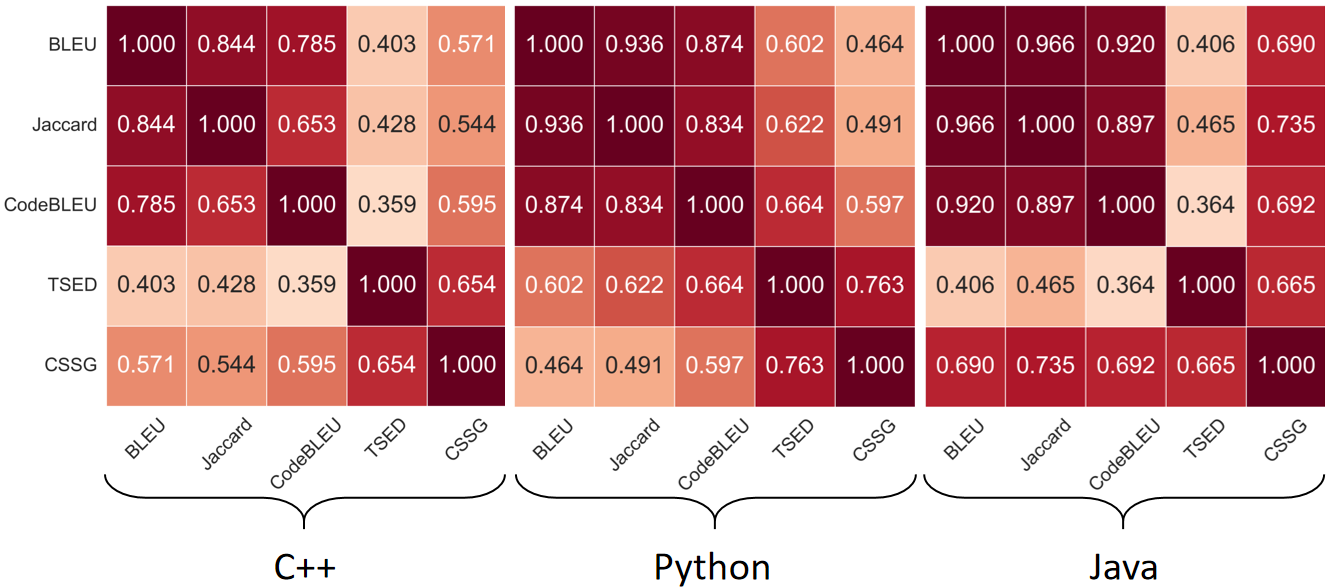}
    \caption{Pearson correlation heatmap among metrics on monolingual code pairs. The correlation patterns suggest that CSSG emphasizes semantic similarities that differ from surface-level metrics while remaining closely related to AST-based approaches.}
    \label{fig:monolingual_pearson}
    \vspace{-10pt}
\end{figure}

Table~\ref{tab:cohen-d-monolingual} reports Cohen's $d$ values for each metric in monolingual settings. CSSG consistently achieves the highest effect size across all three languages, indicating a stronger separation between positive and negative code pairs. This result suggests that CSSG more effectively assigns higher similarity scores to functionally equivalent solutions while maintaining clear distinctions from incorrect implementations within the same language.

To better understand the sources of divergence among different metrics, we compute the Pearson correlations between the Cohen’s $d$ values of each metric, as shown in Figure~\ref{fig:monolingual_pearson}. For both C++ and Python, CSSG exhibits weak correlation with traditional lexical metrics and hybrid approaches, while showing strong correlation with TSED. This trend is consistent with the effect size results. The observed pattern suggests that both CSSG and TSED emphasize structural and logical aspects of code, and that such representations are more effective at capturing functional differences, thereby enabling clearer separation between correct and incorrect implementations.

\subsection{Cross-lingual Similarity Results}
Table~\ref{tab:cohen-d-cross-lingual} summarizes Cohen’s $d$ effect sizes under cross-lingual settings. Compared to the monolingual case, all metrics exhibit reduced discriminative power, reflecting the substantial syntactic and lexical divergence across programming languages. While no single metric dominates every language pair, CSSG achieves the highest average effect size across all cross-lingual settings. Notably, CSSG is the only metric that maintains positive effect sizes for all language pairs, indicating a more stable ability to distinguish positive from negative code pairs under cross-lingual conditions.

\begin{table}[ht]
\centering
\caption{Cohen’s d effect sizes of metrics on cross-lingual code pairs.}
\label{tab:cohen-d-cross-lingual}
\resizebox{\linewidth}{!}{%
\begin{tabular}{lccccc}
\hline
Language Pairs & BLEU & Jaccard & CodeBLEU & TSED & CSSG \\
\hline
(C++, Python) & 0.051 & 0.087 & -0.030 & 0.097 & \textbf{0.131}\\
(C++, Java) & 0.091 & \textbf{0.097} & 0.091 & -0.026 & 0.068\\
(Python, C++) & 0.001 & -0.158 & -0.132 & 0.095 & \textbf{0.174}\\
(Python, Java) & 0.010 & 0.063 & 0.042 & -0.016 & \textbf{0.086} \\
(Java, C++) & -0.091 & -0.117 & -0.149 & \textbf{0.131} & 0.080\\
(Java, Python) & -0.034 & 0.059 & -0.039 & 0.088 & \textbf{0.110} \\
\hline
Average & 0.005 & 0.005 & -0.036 & 0.062 & \textbf{0.108} \\
\hline
\end{tabular}%
}
\end{table}

Lexical metrics such as BLEU and Jaccard show highly unstable behavior, with effect sizes close to zero or even negative in several language directions, underscoring their dependence on surface-level token overlap. CodeBLEU shows modest improvements for some pairs but varies substantially across languages, while TSED performs competitively in limited cases yet remains sensitive to language-specific AST constructions.

In contrast, while CSSG does not dominate every individual language pair, it consistently yields positive effect sizes across diverse cross-lingual settings. This stability arises from its reliance on code semantic graphs, which abstract away language-specific syntax and instead encode data and control dependencies that more directly reflect program semantics. As a result, CSSG is better aligned with functional equivalence across languages, enabling more reliable discrimination between correct and incorrect implementations even when surface forms differ substantially.

\section{Conclusion}

In this work, we proposed CSSG, a code similarity metric based on semantic graph edit distance over enhanced PDGs. By incorporating dependency-aware representations and constrained matching, CSSG captures semantic similarities that are difficult to reflect with lexical or syntax-based metrics.

Experimental results on CodeContests+ show that CSSG more reliably distinguishes correct from incorrect implementations in both monolingual and cross-lingual settings. We further conducted a targeted correlation analysis to examine how different metrics relate to each other, providing insight into why dependency-based similarity leads to more robust discrimination. These findings suggest that CSSG offers a practical semantic alternative for code evaluation beyond surface-level similarity.

\section*{Limitations}

CSSG is proposed as a principled dependency-based similarity metric, and this work focuses on validating its effectiveness rather than exhaustively optimizing all design dimensions. As a result, several limitations should be acknowledged. The performance of CSSG exhibits variation across programming languages and language pairs, reflecting differences in language semantics and the quality of extracted dependency structures. In addition, our graph edit distance formulation adopts uniform costs for all edit operations, which favors simplicity and consistency but may not be optimal for every language or code pattern. These limitations indicate that extrapolating the results to broader settings should be approached with caution and point to directions for further investigation in semantic code similarity modeling.

\section*{Acknowledgments}


\bibliography{custom}
\appendix

\section{Comparison of AST and Semantic Graph Representations}
\label{sec:appendix_comparison}

In this section, we provide concrete examples to illustrate the limitations of Abstract Syntax Tree (AST) representations in capturing program semantics and demonstrate how our proposed Semantic Graph (based on PDGs) overcomes these issues. We focus on three critical aspects of program logic: data flow, control flow, and function call flow.

\subsection{Impact of Data Flow}
\begin{figure}[h]
    \centering
    \vspace{-5pt}
    \includegraphics[width=\linewidth]{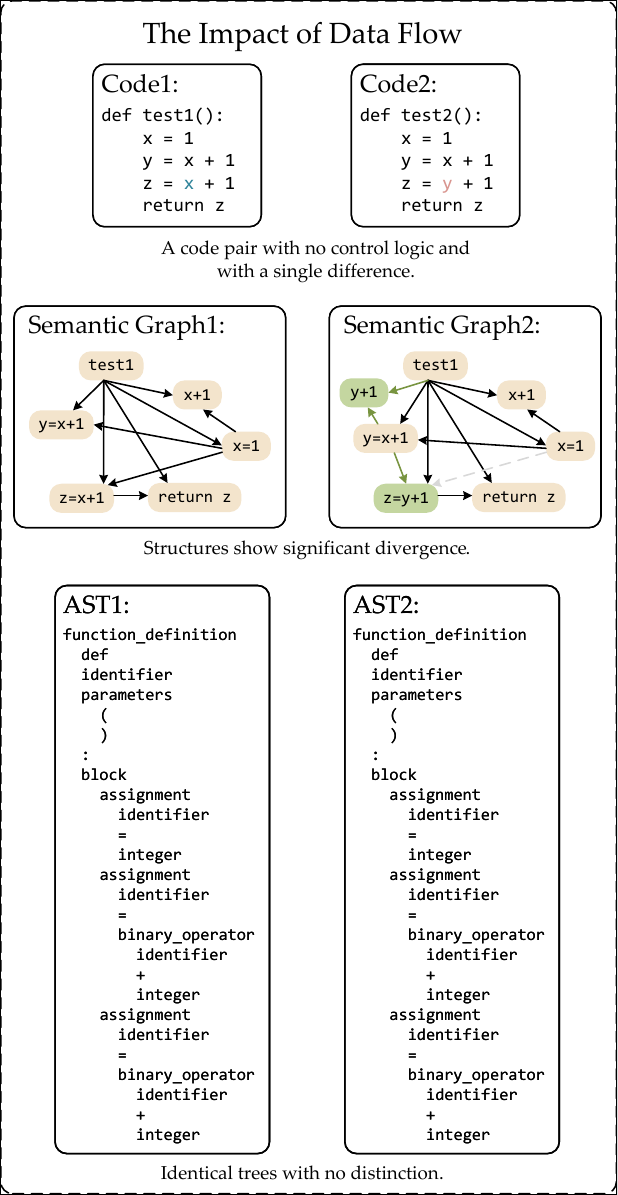}
    \caption{The impact of data flow.}
    \label{fig:CSSG_dataflow}
    \vspace{-10pt}
\end{figure}

\noindent ASTs primarily encode the syntactic hierarchy of code but often fail to distinguish subtle differences in variable usage when the grammatical structure remains unchanged. 
As shown in Figure \ref{fig:CSSG_dataflow}, we consider two code snippets, \texttt{Code1} and \texttt{Code2}, which differ only in the operand used for the assignment of variable \texttt{z} (using \texttt{x} versus \texttt{y}). 
From an AST perspective, both snippets generate identical tree structures consisting of a function definition block followed by a sequence of assignment statements. A similarity metric based purely on AST structure (or tree edit distance without strict identifier matching) would erroneously treat these two programs as identical.
In contrast, the Semantic Graph explicitly models data dependencies. In \texttt{Graph1}, the node for \texttt{z=x+1} has an incoming data dependency edge from \texttt{x=1}, whereas in \texttt{Graph2}, the node \texttt{z=y+1} relies on \texttt{y=x+1}. This topological difference results in a measurable edit distance, correctly reflecting the semantic divergence between the two snippets.

\subsection{Impact of Control Flow}
\begin{figure}[h]
    \centering
    \vspace{-5pt}
    \includegraphics[width=\linewidth]{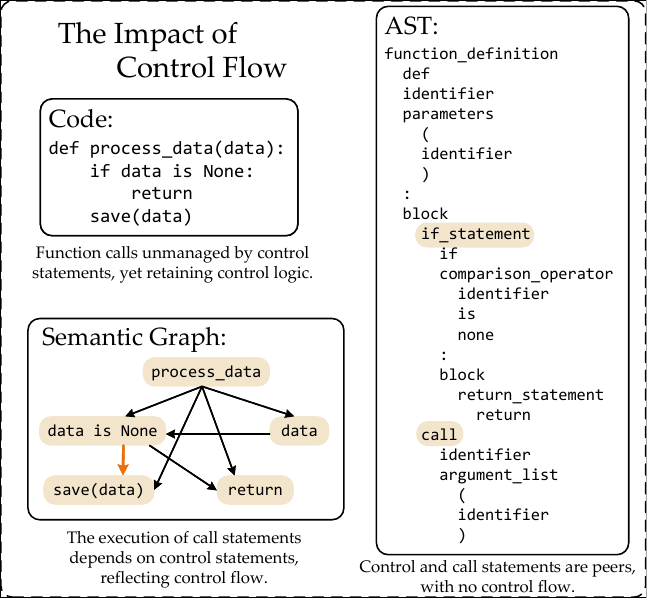}
    \caption{The impact of control flow}
    \label{fig:CSSG_controlflow}
    \vspace{-10pt}
\end{figure}

\noindent ASTs represent statements as sequential siblings within a block, which can obscure the conditional logic governing their execution.
Figure \ref{fig:CSSG_controlflow} illustrates a function \texttt{process\_data} where the execution of \texttt{save(data)} is implicitly guarded by the preceding \texttt{if} statement. In the AST representation, the \texttt{if\_statement} and the \texttt{call} (to \texttt{save}) appear as peer nodes within the function block. The AST structure captures their textual order but fails to explicitly encode the causal relationship—that the call depends on the control predicate evaluating to false.
The Semantic Graph, however, introduces a control dependency edge (highlighted in orange) connecting the predicate \texttt{data is None} to the \texttt{save(data)} node. This edge signifies that the execution of the target node is conditional, thereby embedding the control logic directly into the graph structure.

\subsection{Impact of Function Call Flow}
\begin{figure}[h]
    \centering
    \vspace{-5pt}
    \includegraphics[width=\linewidth]{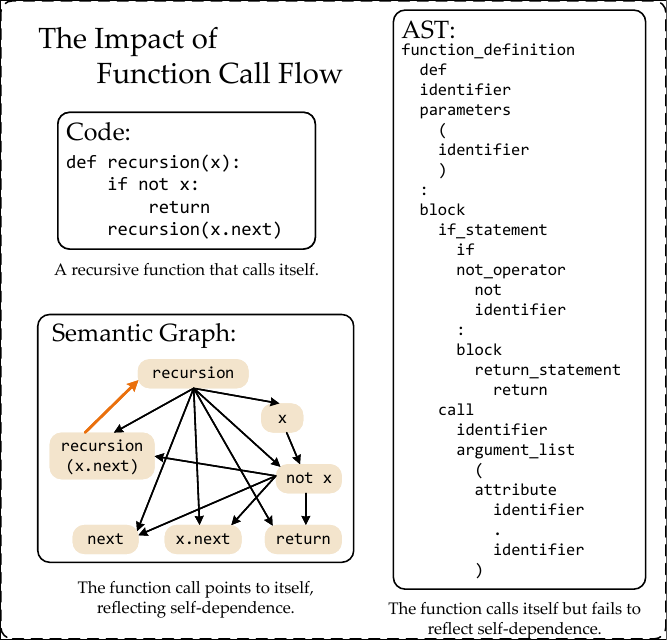}
    \caption{The impact of function call flow}
    \label{fig:CSSG_functioncallflow}
    \vspace{-10pt}
\end{figure}

\noindent Function call flows, such as recursion or inter-procedural calls, are difficult to represent naturally in a strictly hierarchical tree structure.
Figure \ref{fig:CSSG_functioncallflow} demonstrates a recursive function. The AST represents the recursive call simply as a \texttt{call} node nested deeply within the \texttt{if} block. Because ASTs are by definition acyclic trees, they cannot explicitly represent the cycle inherent in recursion; the relationship between the call site and the function entry is lost structurally.
Our Semantic Graph addresses this by introducing function call edges. As seen in the figure, the graph includes a directed edge from the call site \texttt{recursion(x.next)} back to the function entry node \texttt{recursion}. This creates a cycle in the graph, accurately topologically representing the recursive nature of the algorithm and capturing the self-dependence that defines the function's semantics.




\end{document}